\documentclass[aps,prc,twocolumn,amssymb,superscriptaddress,noeprint,floatfix]{revtex4-1}
\newcommand{\rpn}{\textit{r}~process}
\newcommand{\rpa}{\textit{r}-process}

\usepackage[colorlinks,citecolor=blue]{hyperref}
\usepackage[colorinlistoftodos]{todonotes}
\usepackage[normalem]{ulem}
\usepackage{graphicx}
\usepackage{amsmath}
\graphicspath{{./}{figures/}}

\begin{document}

\title{Fission and the \textit{r}-process nucleosynthesis of translead nuclei in
neutron star mergers}

\author{Samuel~A.~Giuliani}
\email{giuliani@nscl.msu.edu}
\affiliation{Department of Physics and Astronomy and NSCL/FRIB Laboratory, Michigan
State University, East Lansing, Michigan 48824, USA}

% ORCID 0000-0002-3825-013
\author{Gabriel~Mart{\'\i}nez-Pinedo}
\email{g.martinez@gsi.de}

\affiliation{GSI Helmholtzzentrum f\"ur
  Schwerionenforschung, Planckstra{\ss}e~1, 64291 Darmstadt, Germany}
\affiliation{Institut f{\"u}r Kernphysik (Theoriezentrum), Department
  of Physics, Technische Universit{\"a}t Darmstadt,
  Schlossgartenstra{\ss}e 2, 64298 Darmstadt, Germany}
\affiliation{Helmholtz Forschungsakademie Hessen f{\"u}r FAIR, GSI
	Helmholtzzentrum f{\"u}r Schwerionenforschung, Planckstra{\ss}e 1, 64291
	Darmstadt, Germany}

\author{Meng-Ru~Wu}
\email{mwu@gate.sinica.edu.tw}
\affiliation{Institute of Physics, Academia Sinica, Taipei, 11529, Taiwan}
\affiliation{Institute of Astronomy and Astrophysics, Academia Sinica, Taipei,
10617, Taiwan}
\affiliation{Physics Division, National Center for Theoretical Sciences, 30013 Hsinchu, Taiwan}

\author{Luis~M.~Robledo}
\email{luis.robledo@uam.es}
\affiliation{Center  for  Computational  Simulation,  Universidad
Polit{\'e}cnica  de  Madrid, Campus  de  Montegancedo, 28660 Madrid,  Spain}
\affiliation{Departamento  de  F{\'i}sica  Te{\'o}rica and CIAFF,  Universidad
Aut{\'o}noma  de Madrid,  28049  Madrid,  Spain}

\begin{abstract}
  We study the impact of fission on the production and destruction of
  translead nuclei during the \rpa\ nucleosynthesis occurring in
  neutron star mergers.  Abundance patterns and rates of nuclear
  energy production are obtained for different ejecta conditions using
  three sets of stellar reaction rates, one of which is based on
  microscopic and consistent calculations of nuclear masses, fission
  barriers and collective inertias. We show that the accumulation of
  fissioning material during the \rpn\ can strongly affect the free
  neutron abundance after the \rpa\ freeze-out.  This leads to a
  significant impact on the abundances of heavy nuclei that undergo
  $\alpha$ decay or spontaneous fission, affecting the radioactive
  energy production by the ejecta at timescales relevant for kilonova
  emission.
\end{abstract}

%% Keywords should appear after the abstract command. 

%%
%% Date and title
\date{\today}%
\maketitle


\section{Introduction\label{sec:intro}}

Sixty years after the seminal works of B$^{2}$FH and
Cameron~\cite{Burbidge1957,Cameron1957}, where the rapid neutron capture process
(\rpn) was firstly indicated as the main mechanism responsible for the
production of the heaviest elements observed in the universe, the GW170817
gravitational wave signal~\cite{Abbott.Abbott.ea:2017} and its associated
AT~2017gfo electromagnetic (EM) counterpart~\cite{Abbott:2017} provided the
first evidence that \rpa\ nucleosynthesis occurs in neutron star mergers
(NSM)~\cite{Lattimer1974,Lattimer1976,Symbalisty1982,Freiburghaus1999}.  This
evidence arose from the observed optical and near-infrared emissions, which were
found to be consistent with a quasi-thermal transient known as kilonova or
macronova powered by the radioactive decay of freshly-synthesized \rpa\
nuclei~\cite{Li1998,Metzger2010,Roberts2011,Goriely2011}.  However, whether NSM
are the main astrophysical site contributing the production of r-process
elements in the Galaxy remains an open question~\cite{Cowan.Sneden.ea:2019}. This
is because despite the large amount of information extracted from the
multimessenger observations, the detailed composition of the ejected material is
still unclear (so far, the only element identified in the ejecta is
Strontium~\cite{Watson.Hansen.ea:2019}). The near-infrared kilonova emission
that was observed at timescales of several days is consistent with predictions
assuming a significant presence of lanthanides (mass fraction $\gtrsim 10^{-2}$)
in the ejecta~\cite{Kasen2017,Kawaguchi:2018ptg}, but the exact range of the
produced nuclei or whether there was a possible presence of heavier elements has
not yet been determined.  In this context, future observations of late-time
($\gtrsim 10$~days) kilonova light curves showing signatures related to the
decay of particular nuclei, together with improved kilonova emission modeling,
would thus provide invaluable information to further progress in our
understanding of the origin of \rpa\
elements~\cite{Zhu2018,Wu2018,Kasliwal.Kasen.ea:2019}.

Both future observations as well as improved theoretical yield predictions are
urgently required.  From the nuclear physics side, one aspect that must be
addressed is the sensitivity of the abundances of long-lived nuclei to nuclear
properties and their impact on kilonova light curves.  The presence of Uranium
and Thorium in metal-poor stars, as well as in the solar system, indicates that
if NSM are a major contributor to the production of \rpa\ nuclei, the \rpa\ path
therein must reach the region of actinides. Therefore, it is likely that fission
happens during and/or after the \rpn.  Previous studies have shown that the
kilonova lightcurves, particularly at late times, depend on the amount of
translead nuclei, e.g., those in the mass range $222\leq A\leq 225$ and $A=254$,
whose yields depend on the adopted nuclear mass
model~\cite{Mendoza-Temis2015,Barnes2016a,Rosswog2017,Wu2018}, and/or the
fission probabilities of heavy nuclei during their decay to
stability~\cite{Zhu2018,Vassh2018,Vassh2019}. However, crucial understanding of
the role played by fission in the production and destruction of translead nuclei
during and after the \rpn\ is still lacking.

In this paper, we study the production of translead nuclei during the \rpa\
nucleosynthesis using three different sets of stellar reaction rates and
trajectories representing three different ejecta conditions in NSM\@.  In
particular, we focus on the role that fission plays in the destruction of very
heavy elements and the implications for the electromagnetic transients powered
by the radioactive decay of the synthesized nuclei. The paper is organized as
follows: Section~\ref{sec:methods} discusses the nuclear properties underlying
the stellar reactions rates and the different trajectories employed in this
work; Section~\ref{sec:results} reports the main results concerning the
evolution of total abundances and nuclear energy release rates; finally,
Section~\ref{sec:conclusions} summarizes the main findings and outline future
works.

\section{Method}\label{sec:methods}

One of the major challenges in \rpa\ nucleosynthesis calculations is to study
the impact of nuclear properties in the abundance patterns and kilonova light
curves. The difficulties arise from the fact that the nuclear reaction network
calculations simulating the \rpa\ nucleosynthesis require the knowledge of
nuclear masses, reactions rates and decay properties for several thousands of
nuclei placed between the valley of stability and the neutron drip-line. Due to
the fact that changes in the nuclear abundances are non-local and that there are
processes like fission that connect very different regions of the nuclear chart,
it is in general very challenging to determine the nuclear properties that
affect the production of particular nuclear species.

Rather critical is the case of fission, where the experimental data suitable for
\rpa\ calculations is particularly scarce. As a result, only few papers so far
addressed the impact of fission during the \rpa\
nucleosynthesis~\cite{Thielemann1983, Panov2004, Panov2008, Petermann2012a,
Korobkin2012, Panov2013a, Goriely2015, Goriely2015a, Eichler2015,
Mendoza-Temis2015, Mumpower2018, Vassh2018, Holmbeck.Sprouse.ea:2019}.  The aim
of this paper is to improve the understanding of the role played by fission in
the production of translead nuclei and its possible relevance for kilonova by
employing a recently developed set of reaction rates and fission properties
based on the Barcelona-Catania-Paris-Madrid (BCPM) energy density functional
(EDF)~\cite{Baldo2013,Giuliani2013}. 

\begin{figure*}[tbh] 
	\includegraphics[width=\textwidth]{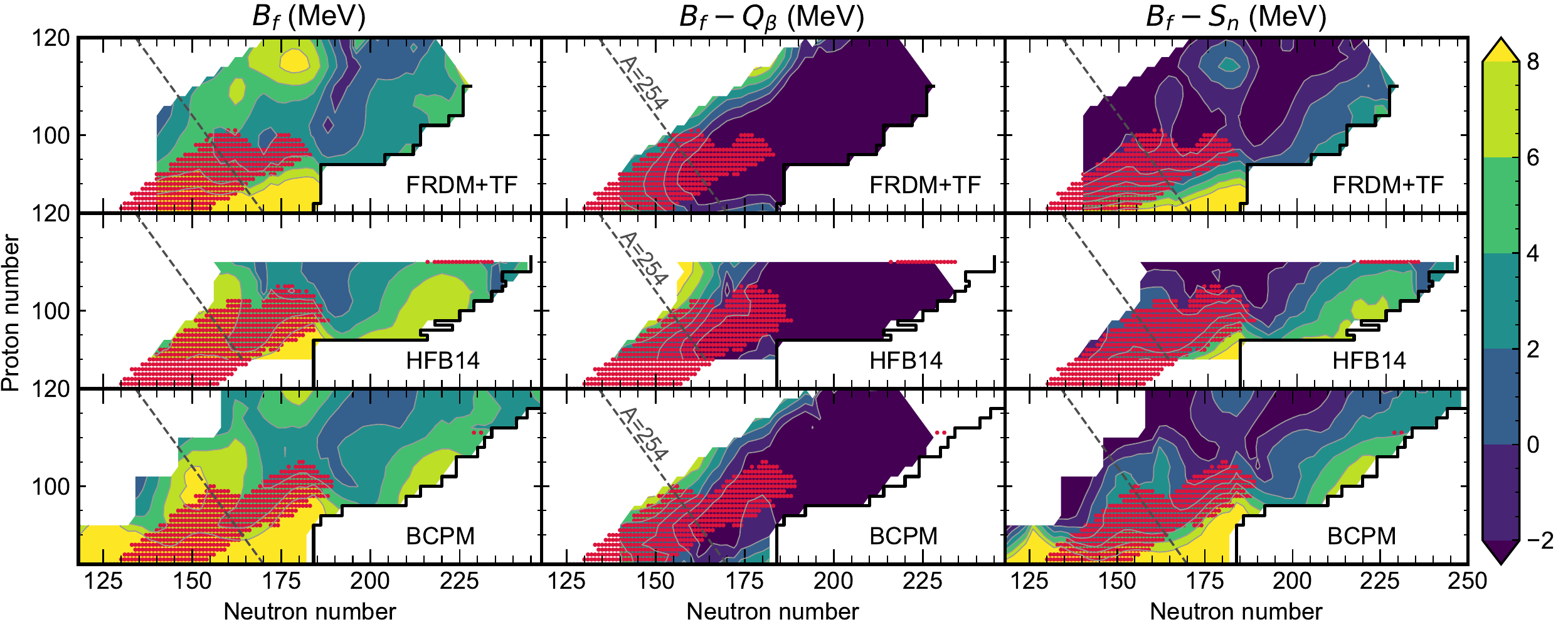}
	\caption{Highest fission barrier ($B_f$), and energy windows for
	$\beta$-delayed fission ($B_f - Q_\beta$) and neutron-induced fission ($B_f
	- S_n$) predicted by FRDM+TF (upper panels), HFB14 (middle panels) and
	BCPM (lower panels) as a function of proton and neutron number. $B_f$
	and $S_n$ values correspond to the nucleus with $Z$ protons and $N$
	neutrons, while $Q_\beta$ values correspond to the FRDM prediction for
	the $(Z-1,N+1)$ parental nucleus. All the quantities are in MeV. Red
	circles indicate the \rpa\ nuclei produced at $t\sim10\,$s in the hot
	dynamical ejecta nucleosynthesis.
	\label{fig:barriers}} 
\end{figure*}

The BCPM neutron-induced reaction rates, $\alpha$-decay rates and spontaneous
fission lifetimes were obtained using the nuclear masses, fission barriers and
collective inertias predicted by the BCPM EDF~\cite{Giuliani2018}.  This is one
of the few attempts to derive a set of reaction rates and nuclear decay
properties suited for \rpa\ calculations from a consistent nuclear input (see
also Ref.~\cite{Goriely2015,Vassh2018,Vassh2019}). Since $\beta$-decay rates are not
available for this functional we employed the finite range droplet model (FRDM)
$\beta$-decay rates~\cite{Moller2003} and derived a set of $\beta$-delayed
fission rates based on BCPM fission barriers and an estimate of the FRDM beta
strength function from the neutron emission probabilities. For nuclei with $Z <
84$, where fission is not expected to play a relevant role, we use the neutron
capture rates based on the FRDM masses~\cite{Rauscher2000} as detailed in
Ref.~\cite{Mendoza-Temis2015}. 

In the literature, two sets of reaction rates have been widely used in \rpa\
calculations involving fission. The rates from Panov \textit{et
al.}~\cite{Panov2010} are based on a combination of FRDM nuclear
masses~\cite{Moller1995} and Thomas-Fermi (TF) fission barriers~\cite{Myers1999}
(see e.g.\ Ref.~\cite{Eichler2015,Mendoza-Temis2015}). More recently, the
Brussels group derived a set of reaction rates~\cite{Xu2013} based on Skyrme-EDF
calculations using the HFB-24 mass model~\cite{Goriely2013b} and the HFB-14
fission properties~\cite{Goriely2009} (see e.g.\
Ref~\cite{Goriely2015,Goriely2015a}). In the following, we will refer to these
two set of reaction rates as FRDM+TF and HFB14, respectively, and we will use
them as reference models to asses the impact of fission properties of translead
nuclei in the \rpa\ nucleosynthesis. This requires the additional calculation of
stellar reaction rates for photodisintegration, that we obtained from the
neutron capture rates by detailed balance.

Fig.~\ref{fig:barriers} shows the FRDM+TF, HFB14 and BCPM predictions of the
highest fission barrier ($B_f$), the difference between fission barrier and
$\beta$-decay $Q$-value ($B_f - Q_{\beta}$)~\footnote{We stress that the BCPM
	and HFB14 $\beta$-delayed fission rates are based on $\beta$-strength
	functions, and hence $Q_{\beta}$ values, predicted by FRDM\@. For
	consistency the lower (middle) middle panel of Fig.~\ref{fig:barriers}
	shows the difference between the BCPM (HFB14) fission barriers and FRDM
$Q_\beta$ values, since the latter were used to determine the maximum
$\beta$-decay energy.} and the difference between fission barrier and neutron
separation energy ($B_f - S_n$) for nuclei with $Z \ge 84$~\footnote{The middle
	row of Fig.~\ref{fig:barriers} shows blank values for nuclei with $84 <
	Z < 90$ because there is no calculation of fission barriers in this
	region using the HFB14 model. In the set of reaction rates derived by
the Brussels group~\cite{Xu2013}, the fission rates for nuclei with $Z < 90$ has
been obtained from ETFSI calculations~\cite{Mamdouh2001} (see~\cite{Koning2007}
for more details)}. These quantities provide a rough estimation of the stability
of each nucleus against the different fission modes: spontaneous fission,
$\beta$-delayed fission and neutron-induced fission, respectively. Evidently,
the smaller these values are, the larger the fission probabilities become.
Looking at nuclei close to the neutron dripline, it is clear that BCPM and HFB14
predict systematically larger fission barriers compared to TF, particularly in
the vicinity of the $N=184$ shell closure. In Section~\ref{sec:results} we will
show how these properties, fission barriers, neutron separation energies and
$Q_\beta$ values, determine the amount of material that can be accumulated in
the heaviest region of the \rpa\ nucleosynthesis.

\begin{figure}[tbh]
  \centering
  \includegraphics[width=0.95\columnwidth]{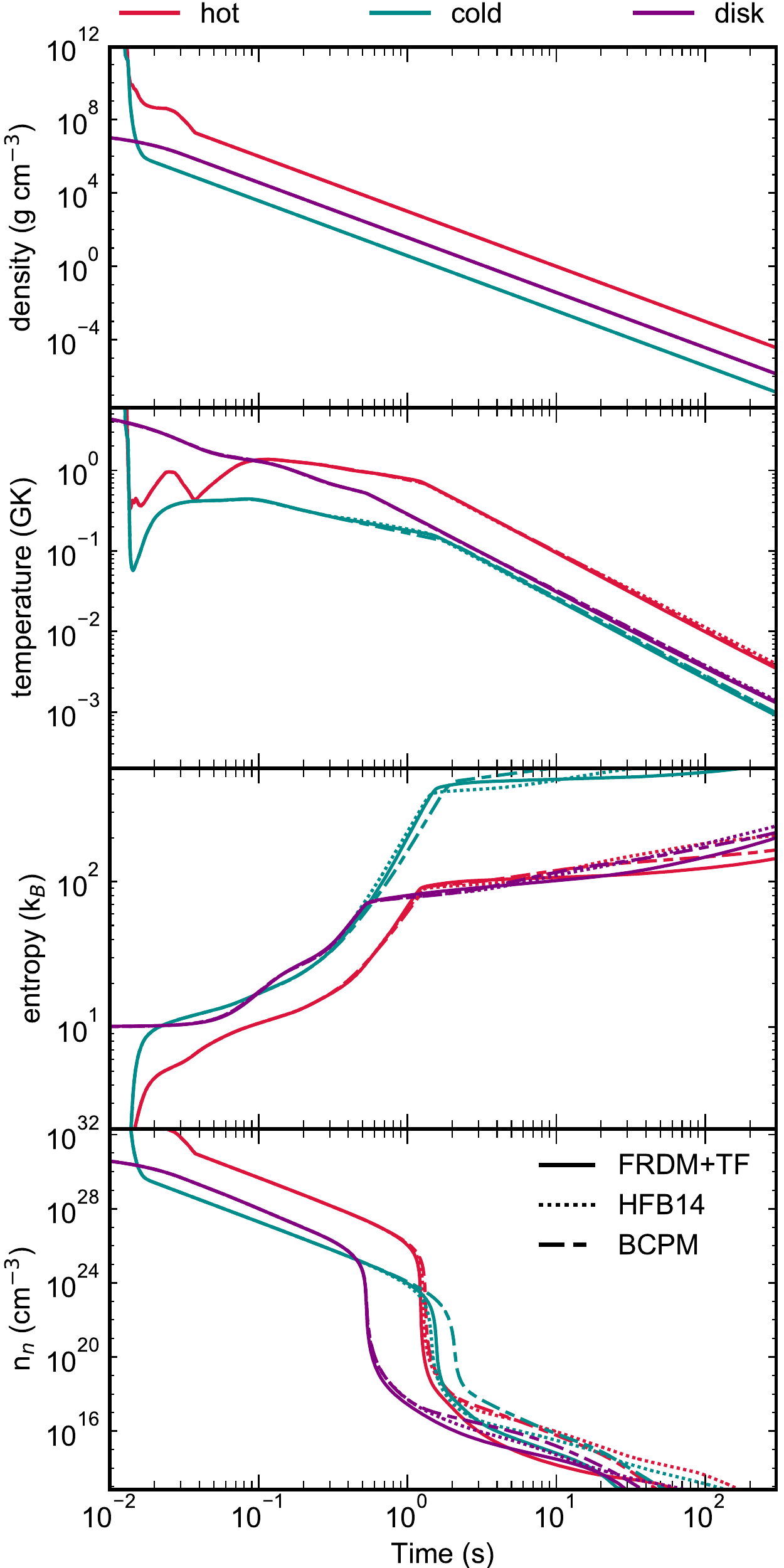} 
  \caption{Evolution of the different thermodynamic variables (from top to
  bottom): mass density, temperature, entropy and free neutron number density
  $n_n$. The different curves represent the predictions obtained with BCPM
  (dashed lines), HFB14 (dotted lines) and FRDM+TF (solid lines) for three
  different trajectories: dynamical hot (red curves), dynamical cold (blue
  curves) and accretion disk (purple curves) (see text for
  details).\label{fig:trajectories}}
\end{figure}

Regarding the astrophysical scenario, we focused our study in the \rpa\
nucleosynthesis occurring in NSM\@. In order to reach conclusions that are
independent of the astrophysical conditions, we employed three trajectories
representing different kind of ejecta conditions.  The evolution of their mass
density, temperature, entropy, and the free neutron number density $n_n$ are
plotted in Fig.~\ref{fig:trajectories}.  The ones labeled by (dynamical) ``hot''
and (dynamical) ``cold'' are trajectories produced by general-relativistic
merger simulation~\cite{Bauswein2013} that were used in previous studies
discussing the role of masses in shaping the \rpa\
abundances~\cite{Mendoza-Temis2015}.  Both of them have initially low entropies
of $\sim 1$~$k_B$ per nucleon and very low electron fraction per nucleon
$Y_e\lesssim 0.05$.  The difference between them is that the dynamical ``hot''
ejecta expand slower than the dynamical ``cold'' one. In the former, the
nuclear energy release during the \rpn\ is able to reheat the ejecta to
temperatures $\gtrsim 1$~GK, while only $\sim 0.2$~GK for the latter~(see Fig.~1
and Eq.~[8] in Ref.~\cite{Mendoza-Temis2015}).  Consequently, an
$(n,\gamma)\rightleftarrows (\gamma,n)$ equilibrium between the neutron-capture
rates and the reverse photo-dissociation rates is only achieved for the former,
but not the latter.  The trajectory labeled ``disk'' is parametrized
following Ref.~\cite{Lippuner2015}, with an early-time expansion timescale
$\tau=10$~ms, initial entropy $s= 10$~$k_B$ per nucleon, and initial
$Y_e=0.15$.  This trajectory mimics the neutron-rich condition found in viscous
outflows from post-merger accretion
disks~\cite{Fernandez2013,Just2015,Wu:2016pnw,Siegel2017,Fernandez:2018kax}.

\begin{figure}[tbh]
	\centering
	\includegraphics[width=0.90\columnwidth]{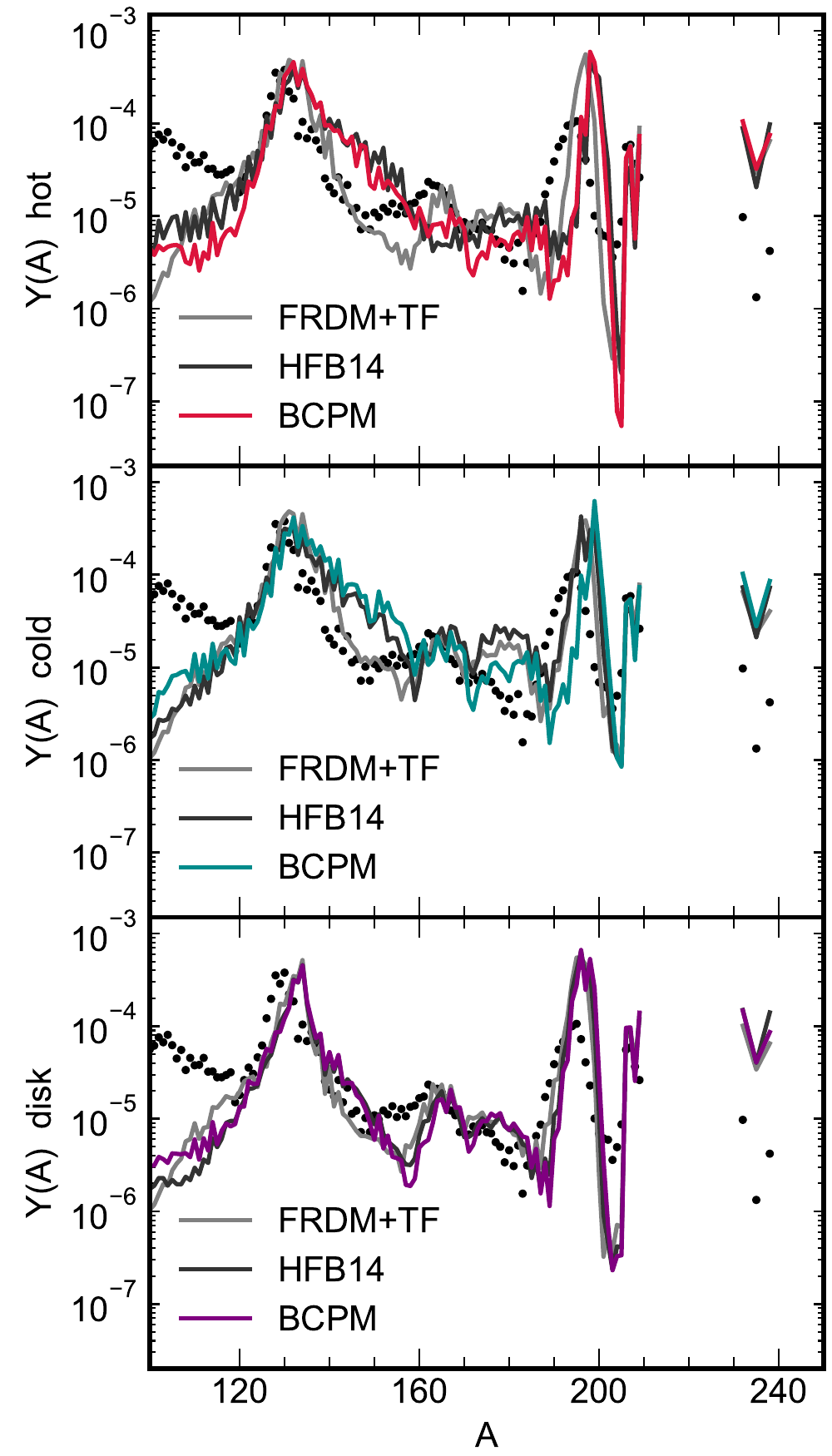}
	\caption{Abundances as functions of mass number predicted by BCPM,
	HFB14 and FRDM+TF at 1 Gyr for the three different ejecta conditions:
	dynamical hot (upper panel), dynamical cold (middle panel) and accretion
	disk (lower panel). As a reference, black dots show the renormalized
	solar \rpa\ abundances.\label{fig:yfinal}}
\end{figure}

Fig.~\ref{fig:yfinal} shows the \rpa\ abundances predicted by FRDM+TF, HFB14 and
BCPM models at the time of $1$~Gyr for these three different ejecta conditions.
All the abundances reproduce the main features of the ``strong'' \rpa\ pattern,
where elements from the second peak up to actinides have been synthesized.
Nevertheless, substantial differences between the predicted abundances are
observed that will be discussed in the next section.

\section{Results\label{sec:results}}

\begin{figure*}[tbh] 
  \centering
  \includegraphics[width=\textwidth]{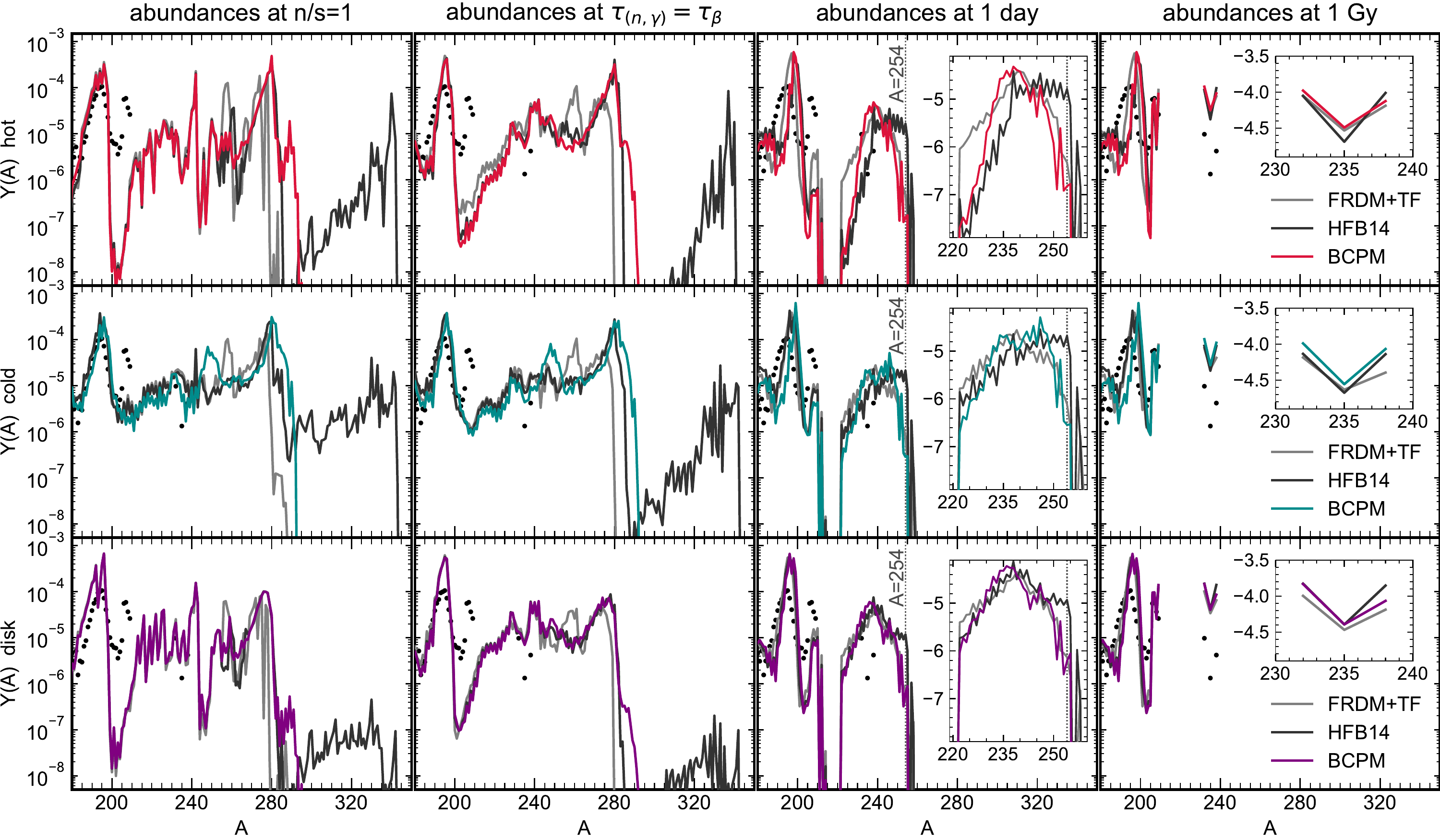}
  \caption{Abundances as functions of mass number predicted by FRDM+TF, HFB14
  and BCPM at four different times. Each row represent a different type of
  ejecta: dynamical hot (upper panels), dynamical cold (middle panels) and
  accretion disk (lower panels). Insets show the abundances prediction in
  $\log_{10}$ scale for particular mass regions. At $t=1$~day, mass number
  $A=254$ is marked with a grey vertical dotted line. Black dots represent
  solar \rpa\ abundances, which are renormalized by the same factor in all
  plots.\label{fig:yscenario}}
\end{figure*}

In order to gain insight into the origin of the differences in abundances shown
in Fig.~\ref{fig:yfinal}, we show in Fig.~\ref{fig:yscenario} the \rpa\
abundances of nuclei beyond $A = 180$ predicted by FRDM+TF, HFB14 and BCPM in
each scenario at four different stages of the evolution: at freeze-out, defined
as the moment when the neutron-to-seed ratio $n/s=1$ (where ``seed'' includes
all nuclei heavier than $^{4}$He); at the moment when the average timescale for
neutron captures $\tau_{(n,\gamma)}$ equals the average timescale for $\beta$
decays $\tau_{\beta}$; at 1~day, which is taken as a timescale indicative for
kilonova observations; and the final abundances calculated at 1~Gyr. For
convenience we group this four time steps in three different stages
characterizing the evolution of the \rpa\ nucleosynthesis: 
\begin{enumerate}
	\item The neutron-capture phase, which begins when the material becomes
		gravitationally unbound and lasts until the freeze-out. During
		this phase, the heaviest region of the nuclear chart is reached
		by successive neutron captures and $\beta$ decays.
	\item The freeze-out phase, which spans the first seconds after
		the freeze-out and during which the average timescale for
		neutron captures $\tau_{(n,\gamma)}$ becomes smaller than the
		average timescale for $\beta$ decays $\tau_{\beta}$. 
	\item The post freeze-out phase, when the material starts to decay
		towards the valley of stability and the abundances pattern is
		shaped to its final distribution shown in Fig.~\ref{fig:yfinal}.
\end{enumerate}

The impact of fission on the \rpa\ nucleosynthesis varies during these three
phases but it mostly manifests through two effects: A direct one,
related to the change in the abundances due to the fission rates and yields; and
an indirect one, induced by the neutron emission of fission fragments (and
the subsequent neutron captures). In the following subsections, we will discuss
how these effects impact the \rpa\ abundances, the evolution of free neutron
densities and the rate of energy production at timescales that are relevant for
kilonova observations.

\subsection{Impact of fission during the \rpn~\label{sec:fissear}}

\begin{figure}[tbh]
	\centering
	\includegraphics[width=0.95\columnwidth]{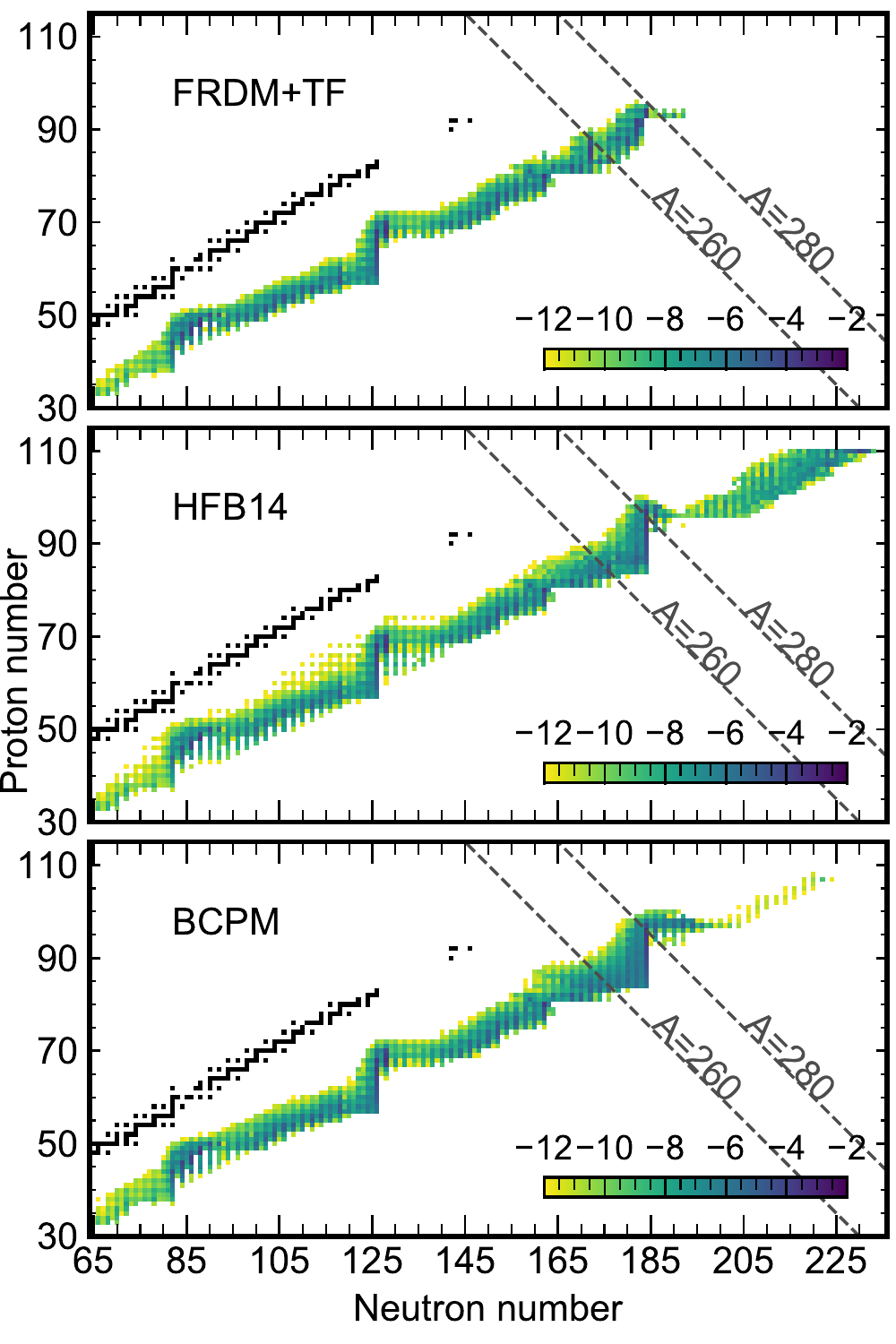}
	\caption{Abundances (in $\log_{10}$ scale) at freeze-out predicted by
	FRDM+TF (upper panel), HFB14 (middle plot) and BCPM (lower plot) in the
	hot dynamical ejecta. Black squares represent stable
	nuclei.\label{fig:nz_plane}}
\end{figure}

We start by determining the mass region in the nuclear chart that is sensitive to
the variations in the physics input described in Section~\ref{sec:methods}.  At
the freeze-out we find that the contribution of $Z \geq 84$ elements in the hot
and accretion scenario is negligible for nuclei with $A \leq 230$ and
constitutes more than 95\% of the $A \geq 252$ abundances. In the cold scenario
these ranges reduce to $A \leq 225$ and $A \geq 246$.  The left column of
Fig~\ref{fig:yscenario} shows that the abundances predicted by the three sets of
reaction rates are visibly different at the freeze-out. In particular, FRDM+TF
exhibits a peak at $A \approx 260$ which is mostly absent in BCPM and HFB14,
while both mean-field models predict a large accumulation of material around $A
\gtrsim 280$. By comparing the nuclear properties of the three models, we found
that two main factors contribute to determine these variations.
First, jumps in the neutron separation energies (and, consequently, in shell
gap energies) can entail accumulation of material at different mass numbers,
particularly during this initial stage of the evolution when neutron-captures
dominate over $\beta$ decays. Second, changes in the fission barriers modify
the survival probability of nuclei and determine the end of the \rpa\ path.  In
the case of the abundances plotted in Fig.~\ref{fig:yscenario}, FRDM predicts a
strong shell gap at $N=172$, which results in the abundances peak at $A \sim
260$. Conversely, the larger fission barriers and shell gap predicted by HFB14
and BCPM at $N = 184$~\cite{Giuliani2018} are responsible for the larger
accumulation of material at $A \sim 280$. These variations are also visible in
Fig.~\ref{fig:nz_plane}, where the \rpa\ path at freeze-out predicted by BCPM,
HFB14 and FRDM+TF for the hot dynamical trajectory is depicted. At $N = 184$,
the \rpa\ path obtained with BCPM and HFB14 models can substantially populate
nuclei up to $^{280}_{\phantom{0}96}$Cm ($t_{1/2}=84$~ms, according to the
$\beta$-decay half-life predictions of Ref.~\cite{Moller2003}), while the
FRDM+TF \rpa\ path accumulates material mostly around
$^{278}_{\hphantom{0}94}$Pu ($t_{1/2}=32$~ms). We recall that around freeze-out
the abundances are to a very good approximation proportional to the
$\beta$-decay lifetimes (see e.g.\ Ref.~\cite{Arcones2011}).

Fig.~\ref{fig:yscenario} also shows that the accumulation of nuclei with $A
\gtrsim 260$ vary with the astrophysical scenario. We find that in the accretion
trajectory the total abundance of fissioning nuclei is a factor two smaller than
the dynamical scenarios and that no fission cycles occur during the \rpn. This
is because the conditions in this trajectory are less neutron rich than in the
dynamical ones: namely, the initial neutron-to-seed ratio is $n/s \sim 120$,
compared to $n/s \sim 600$ and $n/s \sim 1200$ of the hot and cold dynamical
ejecta, respectively. These conditions do not allow the \rpn\ to efficiently
overcome the $N=184$ shell closure, since the number of free neutrons is mostly
depleted when the material reaches the $A \sim 280$ region.

\subsection{Impact of fission at the freeze-out~\label{sec:yn}}

The bottom panel of Fig.~\ref{fig:trajectories} shows that the choice of the
reaction rates substantially change the evolution of $n_n$ after the freeze-out.
In order to understand the origin of such differences, we study the contribution
of individual reaction channels to the change of the neutron abundance $Y_n=n_n
m_u/\rho$, where $m_u$ is the atomic mass unit and $\rho$ is the mass density.
We notice that after the freeze-out and until $\sim 10^3$~seconds, the total
$dY_n / dt$ is much smaller than the $dY_n / dt$ contribution from single
channels, with the exception of $(\gamma, n)$ which quickly becomes negligible.
This suggests that free neutrons are in a condition of quasi-equilibrium for an
extended period of time:
\begin{equation}
	\frac{dY_n}{dt} = 
	\left. \frac{dY_n}{dt} \right\vert_{\text{prod}} - 
	\left. \frac{dY_n}{dt} \right\vert_{\text{abs}}
	\simeq 0 \,,
\end{equation}
during which nuclei efficiently absorb and release neutrons.  Considering a
network formed by neutron captures, photodissociations, $\beta$ decays and
neutron-induced/$\beta$-delayed/spontaneous fission, one gets that the evolution
of $Y_n$ can be written as:
\begin{equation}
  \label{eq:dyndt}
  \begin{split}
	  \frac{dY_n}{dt} & = Y_s \biggl\{
	    \bar{\lambda}_{(\gamma, n)}
	  + \bar{\nu}_{\beta}  \bar{\lambda}_{\beta} 
	  + \bar{\nu}_{\beta\text{fis}}  \bar{\lambda}_{\beta\text{fis}}
	  + \bar{\nu}_\text{sf}  \bar{\lambda}_\text{sf}  \\
	& - Y_n \frac{\rho}{m_u} 
	\Bigl[
	    (1 - \bar{\nu}_{(n, \text{fis})}) \overline{\langle \sigma
              v\rangle}_{(n, \text{fis})} 
	  + \overline{\langle\sigma v\rangle}_{(n, \gamma)} 
          \Bigr]\biggr\}\\
	& \simeq  0\,,
  \end{split}
\end{equation}
where $\bar{\nu}_i$ is the neutron multiplicity of the channel $i$ that releases
neutrons, $\lambda_i$ the channel rate and $\langle \sigma v \rangle_i$ the
cross section averaged over a Maxwell-Boltzmann energy distribution. The bar
over the different quantities denotes average over the composition of seed
nuclei, $Y_s = \sum_j Y_j$. 
\begin{equation}
  \bar{\lambda}_i = \frac{\sum_{j} \lambda_i(j) Y_{j}}{\sum_j
    Y_j}, \quad \bar{\nu}_i = \frac{\sum_{j,k} k \lambda_{i,k}(j) 
    Y_{j}}{\sum_{j,k} \lambda_{i,k}(j) Y_{j}},
\end{equation}
with $\lambda_{i,k}(j)$ the rate for nucleus $j$ to produce $k$ neutrons via the
reaction channel $i$ and $\lambda_i(j) = \sum_k \lambda_{i,k}(j)$.
Eq.~\eqref{eq:dyndt} allows to estimate $Y_n$ given the seed-averaged decay
rates:
\begin{equation}
  \label{eq:ynt}
  	Y_n \approx 
	\frac{%
	    \bar{\lambda}_{(\gamma, n)} 
            + \bar{\nu}_{\beta}  \bar{\lambda}_{\beta} 
	  + \bar{\nu}_{\beta\text{fis}}  \bar{\lambda}_{\beta\text{fis}} 
	  + \bar{\nu}_{\text{sf}}  \bar{\lambda}_{\text{sf}} \\
	}{%
	    \overline{\langle\sigma v\rangle}_{(n, \gamma)} 
	  +(1 - \bar{\nu}_{(n, \text{fis})}) \overline{\langle \sigma
            v\rangle}_{(n, \text{fis})}
	} 
	\frac{m_u}{\rho} \,,
\end{equation}
which shows that  $Y_n$ is directly proportional to the rates producing neutrons
and inversely proportional to the difference between production and absorption
rates involving neutrons as reactants. This implies that $\beta$ decay,
$\beta$-delayed fission, and spontaneous fission contribute differently to $Y_n$
than neutron-induced fission, and that small variations in fission rates can
substantially modify the evolution of neutron abundances if fission is a
relevant source of neutrons.

\begin{figure}[tb] 
  \centering 
  \includegraphics[width=0.5\textwidth]{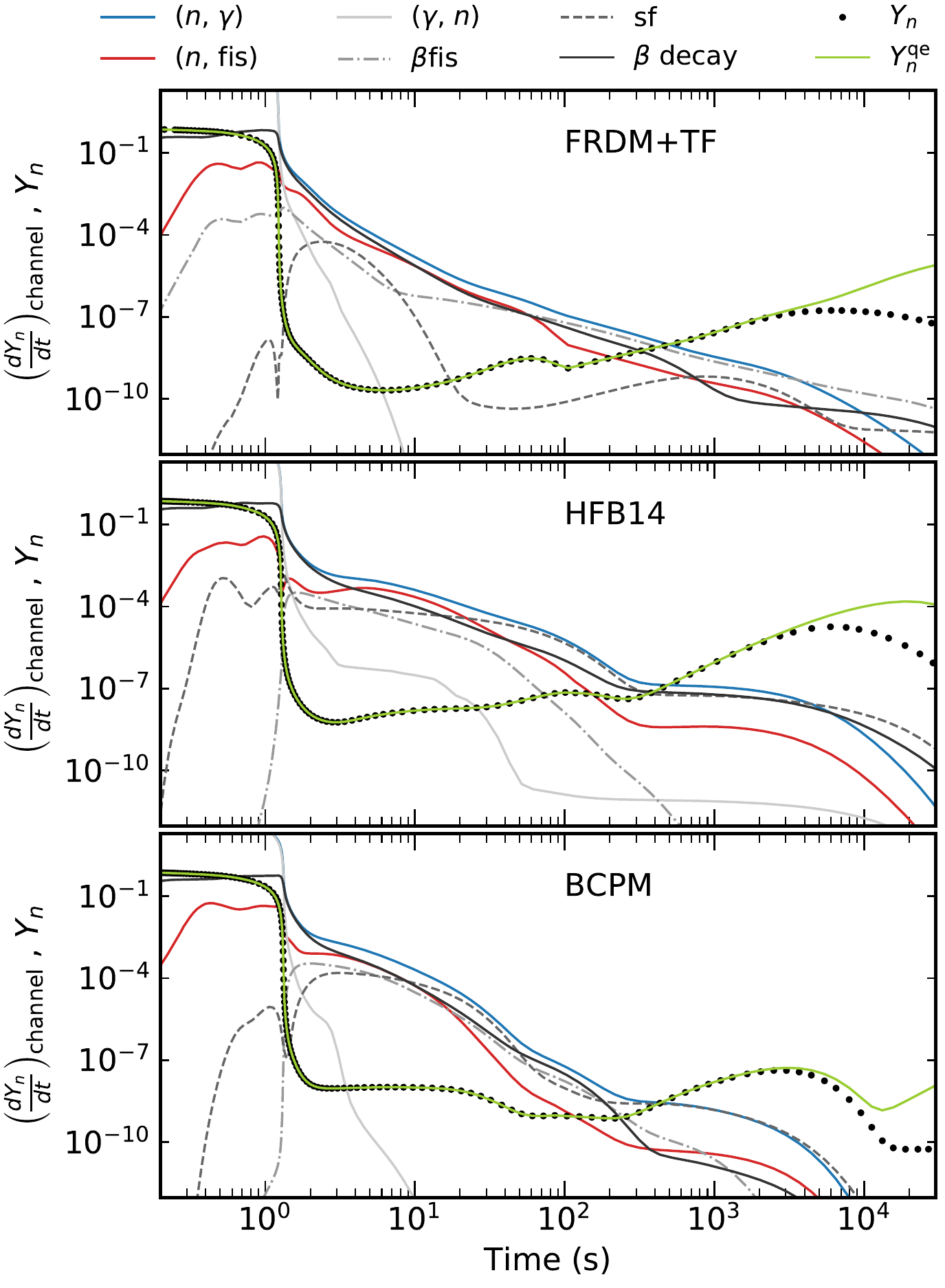}
  \caption{Contribution of single channels in Eq.~\eqref{eq:dyndt} to
  ${dY_n}/{dt}$ obtained with different reaction rates in the hot dynamical
  scenario. The exact neutron abundance $Y_n$ (black circles) are compared with
  those predicted by Eq.~\eqref{eq:ynt} assuming quasi-equilibrium (green
  line).}\label{fig:dyndt}
\end{figure}

To better assess the impact of different decay channels on $Y_n$,
Fig.~\ref{fig:dyndt} shows the individual contributions to $dY_n/dt$ of
Eq.~\eqref{eq:ynt} predicted by different set of reaction rates in the hot
scenario. In this plot one can notice that at 10 seconds the $Y_n$ obtained with
FRDM+TF is more than one order of magnitude smaller than in the HFB14 and BCPM
cases. This is because of the larger $\overline{\langle \sigma v \rangle}_{(n,
\gamma)}$ predicted within FRDM+TF which increases the denominator in
Eq.~\eqref{eq:ynt}, while the accumulation of fissioning nuclei in BCPM and
HFB14 enhances the numerator by boosting the contribution from both fission and
$\beta$-delayed neutron emission. In Sec.~\ref{sec:254Cf} and~\ref{sec:alphaem}
we will discuss how this increase of the free neutron abundances (and,
consequently, of the free neutron number densities) plays an important role in
the production and destruction of nuclei relevant for kilonova observation.

\subsection{Impact of fission on final abundances}

Fig.~\ref{fig:yfinal} shows that the changes in the reaction rates of nuclei
with $Z \geq 84$ in the dynamical scenarios produce large variations in the
final abundances above the second peak ($A \gtrsim 140$) and in the location of
the third peak ($A \sim 195$). The former are directly populated by the fission
fragments of nuclei around $A = 280$, which is a region that the \rpa\ path can
efficiently reach in the case of large fission barriers around $N=184$ (HFB14
and BCPM) within the neutron-rich conditions found in the dynamical scenarios
(see discussion in Sec.~\ref{sec:fissear}). As already explored in different
studies~\cite{Eichler2015, Goriely2015, Goriely2015a, Cote2017, Vassh2018,
Vassh2019}, the final shape in this mass region strongly depends on the
theoretical fission yields assumed for such neutron-rich nuclei (see
Ref.~\cite{Mendoza-Temis2015} regarding the fission fragments distributions used
in this work).  Conversely, the position of the third peak is determined by the
interplay between $\beta$-decay rates and late neutron captures during the
freeze-out~\cite{Eichler2015}. Fig.~\ref{fig:yfinal} shows that the largest
shifts are obtained with HFB14 and BCPM in the hot dynamical scenario, where the
fission of $A=280$ nuclei increases $n_n$ through neutron emission as discussed
in Sec.~\ref{sec:yn}. Fig.~\ref{fig:yfinal} also shows that in the case of the
accretion disk all the models predict a very similar final abundance patterns,
mainly because in this case fission contributes very little to the final
abundance that are determined mostly by the masses and $\beta$-decays of nuclei
with $Z<84$ that remain unchanged in all the calculations.

Finally, we notice that in all the calculations the abundances of nuclei in the
Lead peak ($A \sim 208$) and those of the Uranium and Thorium cosmochronometers
are also very similar (see insets at $t=1$~Gyr in Fig.~\ref{fig:yscenario}).
This result indicates that the progenitors of these nuclei have the same nuclear
properties, which is consistent with our discussion in Sec.~\ref{sec:fissear}
where we argued that modifying the nuclear properties of elements with $Z \geq
84 $ only captures the sensitivity of nuclei with $A \gtrsim 252$. It is
therefore possible to conclude that most the material with $Z \geq 84$ created
during the \rpa\ nucleosynthesis fissions.

\subsection{Fission and the destruction of $A\gtrsim 250$
nuclei~\label{sec:254Cf}}

One feature shown in Fig.~\ref{fig:yscenario} is that, at the time of a day,
BCPM and FRDM+TF predict a drastic drop of the abundances for nuclei with $A
\geq 250$, while in HFB14 calculation this dip is displaced to $A \geq 255$.
These differences have important consequences in terms of kilonova observation,
since the decay by spontaneous fission of $^{254}$Cf, $t_{1/2}= 60.5\pm
0.2$~days~\cite{Phillips1963a}, can sensibly impact the shape and magnitude of
the kilonova lightcurves at $t\gtrsim 100$~days~\cite{Zhu2018,Wu2018}.
Therefore, it is important to understand the mechanisms that are responsible for
the destruction of these nuclei as they also determine the amount of $^{254}$Cf
that survives at kilonova times. For this purpose, we performed additional
calculations by switching off different fission channels (neutron-induced,
$\beta$-delayed and spontaneous fission) and compare the impact of each channel
on the remaining abundance of $A\gtrsim 250$ nuclei at 1 day. While we only
discuss below the results obtained in the hot dynamical case, we note that
similar outcomes are obtained in the cold and accretion disk scenarios.

\begin{figure*}[tbh] 
  \centering 
  \includegraphics[width=0.98\textwidth]{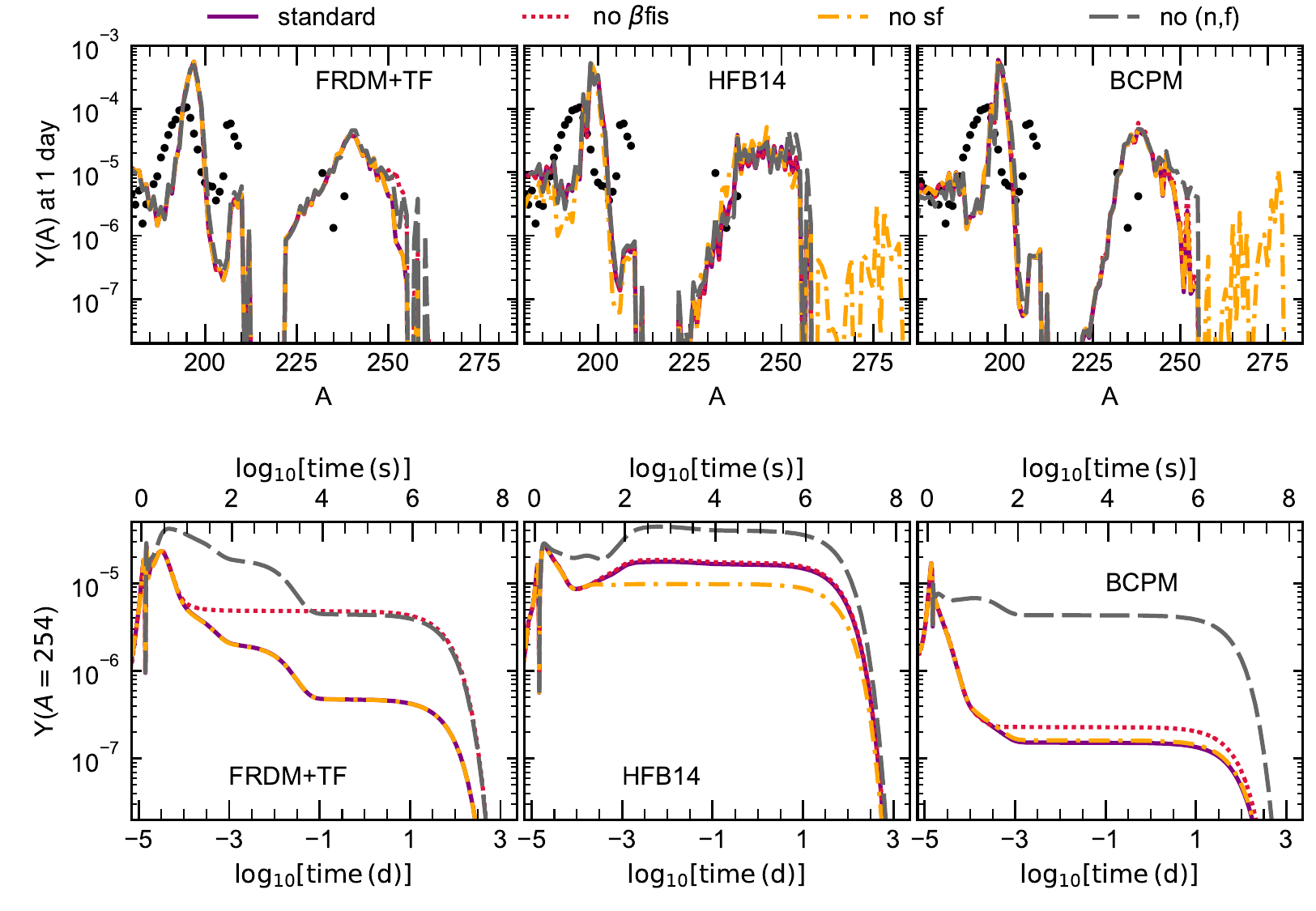}
  \caption{Impact of single fission channels on the abundances of nuclei with
  $A\gtrsim 250$ at 1~day (top panels) and on the evolution of $A=254$
  abundances (bottom panels) for the hot dynamical ejecta. The different curves
  show the abundances predicted when different fission channels are suppressed:
  $\beta$-delayed fission ($\beta$\emph{fis}, red dashed line), spontaneous
  fission (\emph{sf}, yellow dashed line) and neutron-induced fission ($(n,f)$,
  grey dashed line). The purple solid line corresponds to standard calculations,
  when all the fission are included. Left, middle and right panels correspond to
  FRDM+TF, HFB14 and BCPM, respectively.\label{fig:y254}}
\end{figure*}

The upper panels in Fig.~\ref{fig:y254} show the abundances predicted with
FRDM+TF (left panel) HFB14 (middle panel) and BCPM (right panel) when different
fission channels are turned off. Spontaneous and $\beta$-delayed fission are
suppressed from the beginning of the simulation, while neutron-induced fission
was turned off only after the freeze-out. We point out that only theoretical
rates have been switched off. We find that within each set of reaction rates the
abundances drop for the same value of $A$, indicating that the drop is a generic
feature of the behavior of fission barriers in the region. This is confirmed by
comparing the abundance distribution after the \rpa\ freeze-out to the fission
barriers shown in Fig.~\ref{fig:barriers}. Nuclei that are present at $t \sim
10\,\,$s in the hot dynamical ejecta scenario are plotted as solid symbols. Left
panels shows that the abundance distributions closely follows contour lines of
constant fission barrier height, and that none of the models predict the
synthesis of nuclei with $B_f < 2~$MeV. We conclude therefore that the
destruction mechanism is related to the presence of low fission barriers, which
in turn make those nuclei unstable against fission regardless of the
astrophysical environment. For BCPM and FRDM+TF such region inhibits the
survival of nuclei with $A > 250$, while the larger fission barriers predicted
by HFB14 allow nuclei up to $A = 255$ to remain. The only calculations where
nuclei with $A > 255$ survive at 1~day are for the BCPM and HFB14 rates without
spontaneous fission. In both cases, the larger $B_f-Q_{\beta}$ predicted by
these two models around $N=184$ allow part of the material to undergo multiple
$\beta$ decays before entering in the region of low $B_f$ and fission.  One
should notice that in Fig.~\ref{fig:barriers} there are nuclei with negative
energy window for $\beta$-delayed fission $(B_f-Q_\beta)$ that are populated.
The reason for this is twofold: First, the $\beta$-decay proceeds mainly via
states with low excitation energy, hence it is the magnitude of the barrier and
not necessarily $B_f-Q_\beta$ that determines the fission survival probability
after $\beta$-decay. Second, nuclei populated in Fig.~\ref{fig:barriers} have
$S_n < B_f$, as evinced by the right panels in the same plot, which favors the
$(n,\gamma)$ reaction over $(n,\text{fission})$. 

\begin{figure*}[htb] 
  \centering 
  \includegraphics[width=\textwidth]{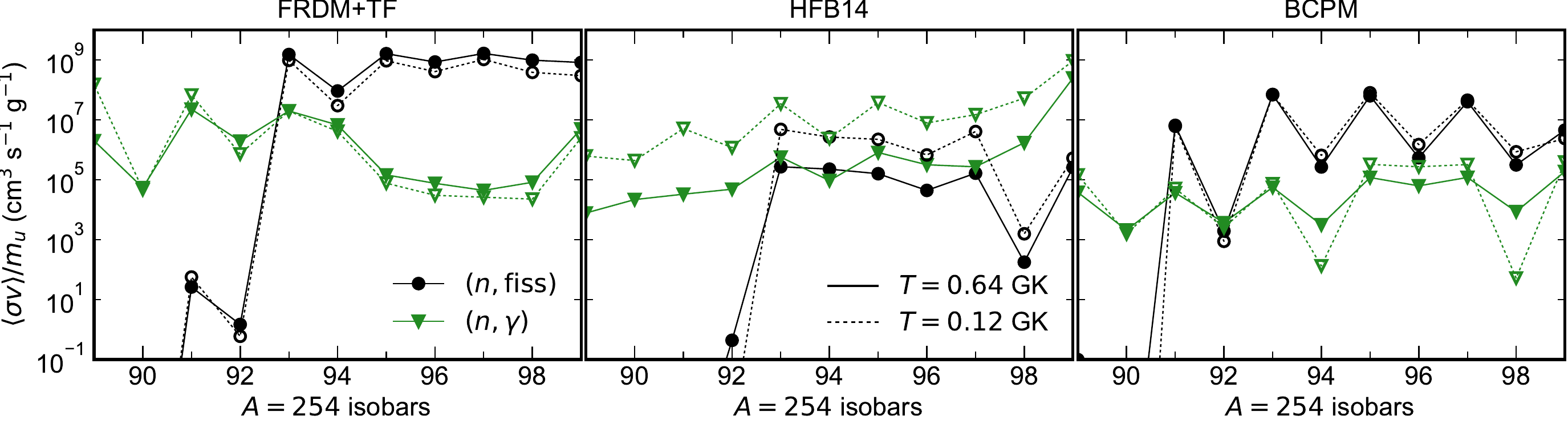}
  \caption{Neutron-induced fission (black circles) and neutron capture (green
	  triangles) stellar reaction rates predicted at $T = 0.64$~GK
	  (solid lines, solid symbols) and $T =
	  0,12$~GK (dotted lines, hollow symbols) by FRDM+TF (left
  panel), HFB14 (middle panel) and BCPM (right panel).}\label{fig:rates}
\end{figure*}

Bottom panels in Fig.~\ref{fig:y254} show the total $A=254$ abundance predicted
in the hot dynamical ejecta by the different sets of reaction rates. From this
plot it is possible to conclude that in all the models neutron-induced fission
is the main mechanism responsible for the destruction of $A=254$ isobars, but
the impact of this channel strongly depends on the adopted nuclear input.  There
are mainly two reasons causing such variations: The first one is related to the
ratio between neutron captures and neutron-induced fission, which determine the
survival probability of the nucleus after capturing a neutron.
Fig.~\ref{fig:rates} shows the neutron capture and neutron-induced fission
reaction rates for different $A = 254$ isobars predicted by the three nuclear
models at 0.64~GK and 0.12~GK, which are the temperatures at freeze-out for the
hot and dynamical trajectory, respectively. In the case of BCPM (FRDM+TF),
neutron-induced fission dominates over neutron captures for nuclei above
$^{254}_{\phantom{0}91}$Pa$_{163}$ ($^{254}_{\phantom{0}93}$Np$_{161}$),
suggesting a more efficient destruction of $^{254}$Cf progenitors compared to
HFB14, for which neutron-induced fission is mostly subdominant along the
isobars. The second aspect are variations in the neutron number density ($n_n$),
which regulate the competition between neutron-induced fission and other decay
channels such as $\beta$-decay rates.  As discussed in Sec.~\ref{sec:yn}, after
the freeze-out BCPM and HFB14 predict larger $n_n$ than FRDM+TF because of the
larger amount of fissioning nuclei emitting neutrons, which in turn boosts the
destruction of $A=254$ isobars in BCPM due to the dominance of neutron-induced
fission over neutron capture showed in Fig.~\ref{fig:rates}. This feature is
reminiscent of the self-sustained mechanism occurring in nuclear reactors, where
the free neutrons released after the freeze-out generate new neutron-induced
fission events. We conclude that the destruction of $A = 254$ isobars should be
considered as a more general feature, where the destruction rate of nuclei post
freeze-out is strongly related to variations in the neutron number density since
this directly determines the neutron captures and neutron-induced fission rates.
On the other hand, Fig.~\ref{fig:y254} shows that $\beta$-delayed fission only
operates to destruct the $A = 254$ isobars at later times with FRDM+TF model,
since the large $B_f-Q_\beta$ predicted by BCPM and HFB14 disfavour this fission
channel (see middle panels in Fig~\ref{fig:barriers}). Finally, in
Fig.~\ref{fig:rates} we observe that the competition between reaction rates is
not affected by variations in temperature given by the different trajectories
explored in this study.

\subsection{Impact of fission on abundance of nuclei with
$A=222-225$~\label{sec:alphaem}}

\begin{figure}[tbh]
	\centering
	\includegraphics[width=0.95\columnwidth]{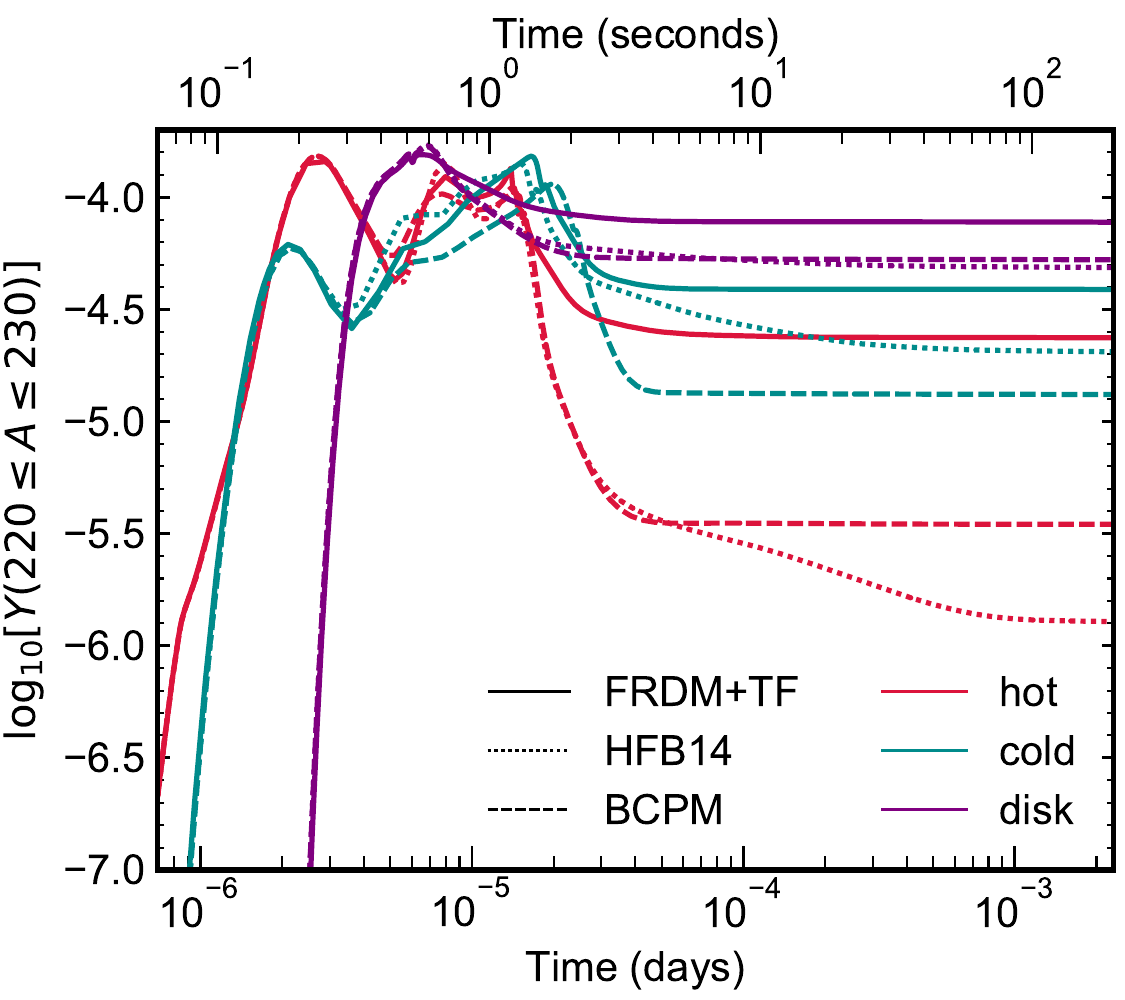}
	\caption{Evolution of the total $220 \le A \le 230$ abundance predicted
	by FRDM+TF (solid lines), HFB14 (dotted lines) and BCPM (dashed lines)
for different trajectories.\label{fig:y220230}}
\end{figure}

Besides $^{254}$Cf, another relevant region for kilonova observations are
actinides with mass number $A=222-225$~\cite{Wu2018}. If these $\alpha$ emitters
can be produced in a substantial amount, their released energy could dominate
the heating rates at timescales of weeks to months, providing a unique signature
of production of heavy nuclei during the \rpn. The insets of
Fig.~\ref{fig:yscenario} at 1~day show that the abundances of such nuclei depend
on the set of stellar reaction rates.  Particularly, BCPM and HFB14 models in
the dynamical scenarios predict smaller abundance compared to FRDM+TF, as shown
in Fig.~\ref{fig:y220230} where the abundance evolution of nuclei with $220 \leq
A \leq 230$ is plotted for the different nuclear models and trajectories. Since
$A=220-230$ abundances are dominated by elements with $Z \leq 83$, whose nuclear
properties are fully determined by FRDM, we conclude that the variations
observed in Fig.~\ref{fig:yscenario} and~\ref{fig:y220230} are driven by the
changes in free neutron number densities discussed in Sec.~\ref{sec:yn}.  This
conclusion is consistent with the fact that the depletion observed in
Fig.~\ref{fig:y220230} occurs in the first 1--2.5 seconds after the freeze-out,
that is the timescale when fissioning nuclei enhance the neutron abundance. As a
consequence, BCPM and HFB14 in the dynamical scenarios show a larger exhaustion
compared to FRDM+TF due to the larger accumulation of fissioning material, while
all the models predict similar final abundances in the accretion case. We point
out that at 1~day the total abundance of nuclei with $220 \leq A \leq 260$ is
similar in all the calculations, suggesting that the underproduction of $\alpha$
emitters in BCPM and HFB14 calculations is related to a transport of material to
heavier masses rather than to a destruction of those nuclei.

\subsection{Heating rates and kilonova light curves}

\begin{figure*}[tb] 
	\centering
	\includegraphics[width=\textwidth]{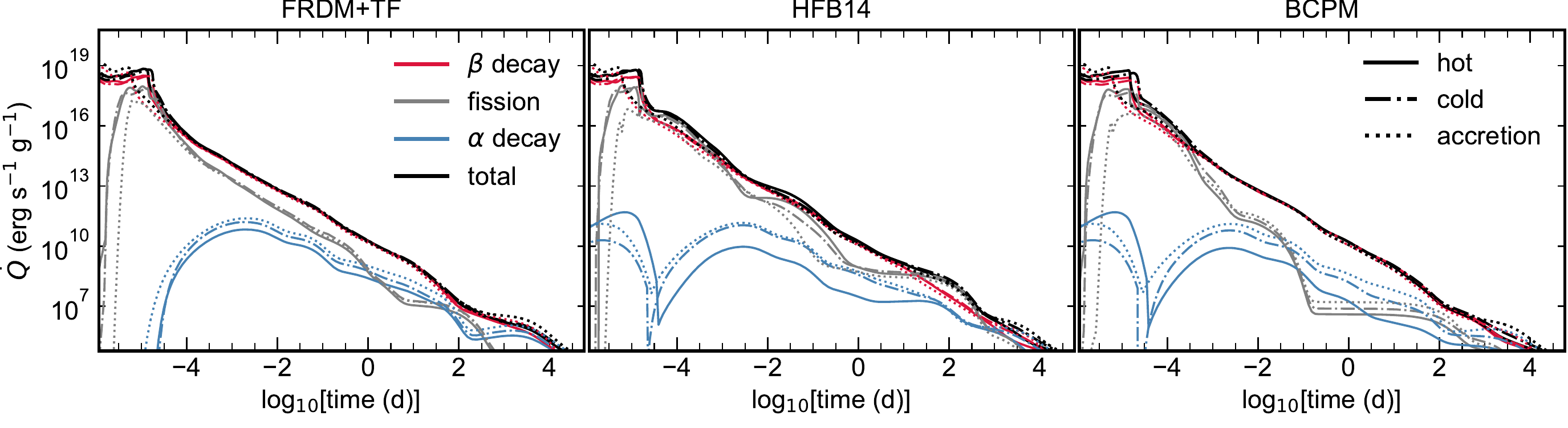} 
	\caption{Radioactive energy emitted by $\beta$ decay (grey lines),
	$\alpha$ decay (blue lines) and fission (red lines) as a function of
	time for different ejecta conditions: dynamical hot (solid lines),
	dynamical cold (dash-dotted lines) and accretion disk (dotted lines).
	Upper, middle and lower panel shows the results predicted by FRDM+TF,
	HFB14 and BCPM, respectively.\label{fig:heating}.}
\end{figure*}

In the previous two sections we analyzed the role of fission in the production
of nuclei that may produce signatures relevant to the synthesis of heavy nuclei
during the \rpn. In order to properly address the possible implications for
late-time kilonova nebular observations, in this Section we discuss the impact
of our calculations in both the nuclear energy release and the ejecta heating
rates. Fig.~\ref{fig:heating} shows the nuclear energy release rate produced by
fission, $\beta$ and $\alpha$ decays as a function of time obtained with
FRDM+TF, HFB14 and BCPM models for the different ejecta conditions. The fission
heating rates in the range 10--1000 days are dominated by the spontaneous
fission decay of $^{254}$Cf. Depending on the set of reaction rates, the
abundances of this nucleus change substantially and hence its contribution to
the nuclear energy release.  For both FRDM+TF and BCPM the contribution of
$^{254}$Cf to the total heating is negligible independently of the astrophysical
scenario considered. For HFB14 heating by $^{254}$Cf dominates around 100~days.
This is due to the fact that for FRDM+TF and BCPM $(n,\text{fiss})$ completely
dominates over $(n,\gamma)$ for nuclei around $A\approx 254$ (see
Figure~\ref{fig:rates} and discussion in Sec.~\ref{sec:254Cf}) while for
HFB14 $(n,\gamma)$ and $(n,\text{fiss})$ are of similar magnitude leading to a
larger abundance of $^{254}$Cf in the three scenarios.

Besides the variations at 10--1000 days, Fig.~\ref{fig:heating} shows that
the models predict different rates also at $t \sim 0.1$~days. Comparing FRDM+TF
and BCPM, one can notice that the latter shows a sharp transition which is
mostly absent in the former. We find that in FRDM+TF the contribution to the
energy production of fission is mainly sustained by the neutron-induced fission
flows (defined as the product of fission rate and nuclear abundance) of
$^{241,242,244}$Pu. In BCPM, such Plutonium isotopes are less abundant and have
smaller neutron-induced fission rates resulting in a quench of the radioactive
energy emitted by fission.  HFB14 predicts a large contribution of fission to
the energy production at timescales of 0.1~days that can even dominate over
$\beta$ decay. This may be an artifact caused by the accumulation of nuclei with
$Z=110$ at the edge of our nuclear network (see middle panels in
Fig.~\ref{fig:barriers}) and will require further calculations with an extended
network.

In addition to the heating from fission and the role of spontaneous fission of
$^{254}$Cf, Fig.~\ref{fig:heating} shows also the contribution of $\alpha$-decay
powered by the decay chains of nuclei with $222 \le A \le 225$ during the
relevant kilonova timescale of 3--100 days~\cite{Wu2018}. We find that for all
models and astrophysical scenarios the contribution of $\alpha$ decay is
subdominant. We shall recall however that the method followed in this work does
not fully capture the nuclear sensitivity of this mass region, since the
progenitors of $222 \le A \le 225$ are elements with $Z \leq 83$. Therefore in
order to better asses the uncertainty in the production of these
$\alpha$-emitters, additional sensitivity studies including variations in the
nuclear properties of lighter elements are in order.
  
\begin{figure}[tbh]
	\centering
	\includegraphics[width=0.95\columnwidth]{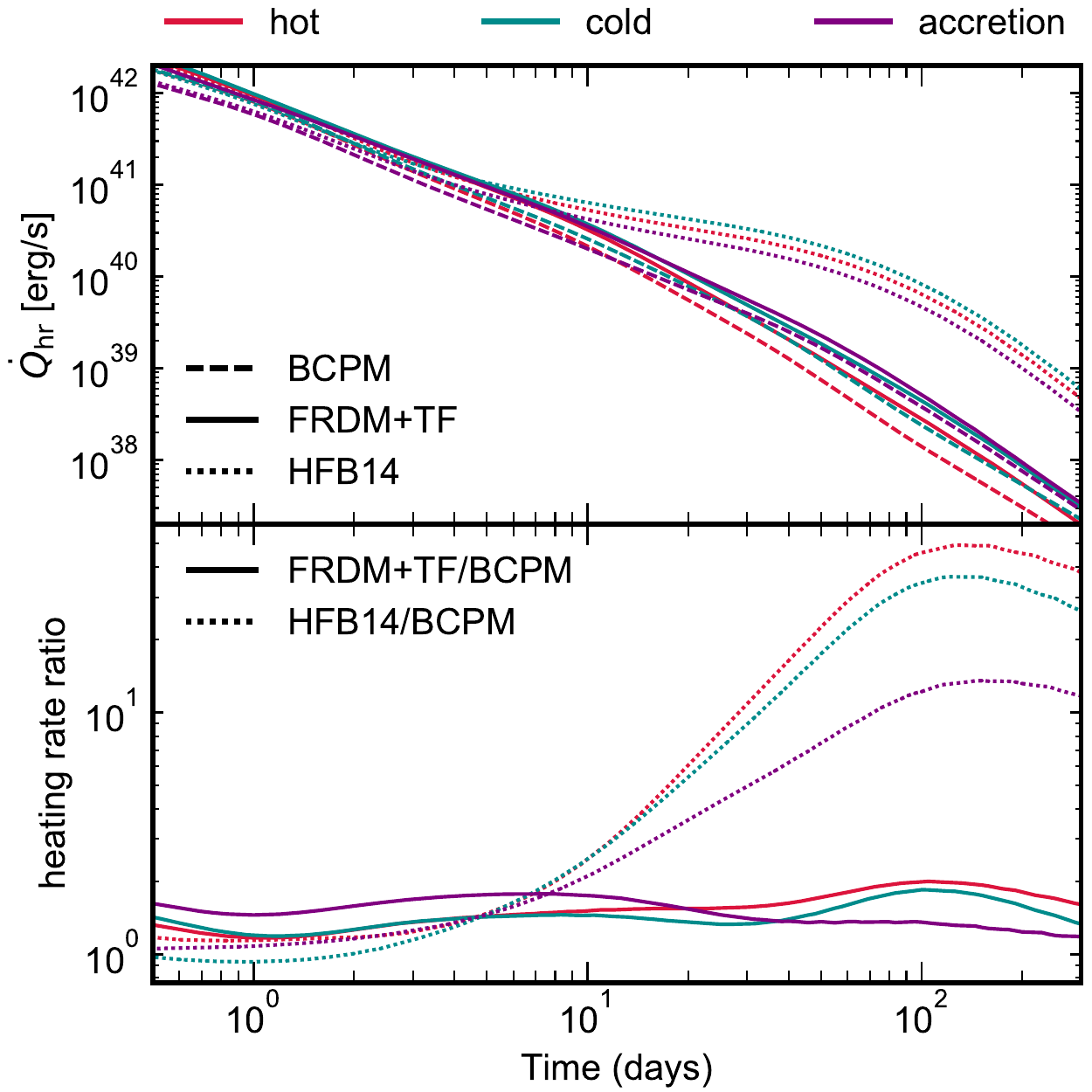}
	\caption{Upper panel: Ejecta heating rate as a function of time
	predicted with BCPM (dash lines), HFB14 (dotted lines) and FRDM+TF
	(solid lines) for different scenarios. Lower panel: Evolution of the
	ratio between ejecta heating rates for different scenarios: FRDM+TF/BCPM
	(solid lines) and HFB14/BCPM (dotted lines).\label{fig:lightcurve}}
\end{figure}

Finally, Fig.~\ref{fig:lightcurve} shows the time evolution of the ejecta
heating rate $\dot{Q}_{hr}$, which mimics the bolometric luminosity of the
kilonova at $t\gtrsim$~days post the lightcurve peak. Calculations were
performed as described in Ref.~\cite{Wu2018}, which include thermalization
corrections, assuming an ejecta mass $M_{ej} = 0.04$~M$_\odot$ with an expanding
velocity $v_{ej} = 0.1 \,\, c$. The impact of the $^{254}$Cf is clearly
noticeable at $t \sim 100\,\,$days, where the predictions obtained with FRDM+TF,
HFB14 and BCPM visibly differ. In the dynamical scenarios, the smaller amount of
$^{254}$Cf predicted with BCPM [$Y(^{254}$Cf$)\approx 1.5\times 10^{-7}$]
results in a heating rate fifty times smaller than the one predicted with HFB14
[$Y(^{254}$Cf$)\approx 1.6\times 10^{-5}$] and two times smaller than the one
obtained with the FRDM+TF rates [$Y(^{254}$Cf$)\approx 4.7\times 10^{-7}$].  One
shall notice that changes in $^{254}$Cf abundances and ejecta heating rates are
not proportional due to the $\beta$-decay contribution to the heating rate.
Nevertheless, this result confirms the high sensitivity of kilonova light curve
to the amount of $^{254}$Cf fissioning at timescales relevant for astronomical
observations~\cite{Zhu2018,Wu2018}, which translates into a large uncertainty in
the heating rates due to variations in the fission properties of translead
nuclei. We also notice that despite the abundances of nuclei with $135 \lesssim
A \lesssim 200$ are largely affected by the direct impact of fission, the
$\beta$-decay heating rate at earlier times are only affected by $\lesssim
50\%$.

\section{Conclusions\label{sec:conclusions}}

We explored the impact of fission on the \rpa\ nucleosynthesis yields
in neutron star mergers and the associated nuclear energy release
rates relevant for kilonovae.  We used three different sets of stellar
reaction rates, one of which was recently developed using consistent
nuclear energy density functional calculations of nuclear masses,
fission barriers and collective inertias~\cite{Giuliani2018}. Our
calculations show that for the most neutron rich conditions, like
those found in the dynamical ejecta, the stability against fission of
nuclei around the neutron shell closure $N=184$ is crucial for the
build-up of fissioning material during the \rpn. The fission of these
material after the \rpa\ freeze-out can release a large amount of
neutrons and significantly alter the free neutrons number density of
the ejecta around 1~s $\lesssim t \lesssim 10$~s.  Consequently, the
neutron-induced fission and neutron-captures associated with these
free neutrons can have strong impact on the abundances of nuclei in
the mass number region $A=220-260$, including the $\alpha$-decaying
nuclei with $222\leq A \leq 225$ and the $^{254}$Cf fission, affecting
the ejecta heating rates on timescales relevant for kilonova light
curve predictions.

In conclusion, we find a connection between the amount of material produced
around $A = 280$ at early stages of the evolution and the amount of $^{254}$Cf
produced at timescales relevant for kilonova observation.  This result suggests
that future detection or non-detection of $^{254}$Cf on kilonova light curves
may help to constraint the yields of nuclei around $A \sim 280$ and learn about
the nuclear properties in a region that in the foreseeable future will not be
experimentally accessed~\cite{Horowitz2018}.

\begin{acknowledgments}
	SAG would like to thank M.~Eichler, S.~Goriely and N.~Vassh for
	insightful discussions.  SAG acknowledges support from the U.S.
	Department of Energy under Award Number DOE-DE-NA0002847 (NNSA, the
	Stewardship Science Academic Alliances program). The work of GMP is
	partly supported by the Deutsche Forschungsgemeinschaft (DFG, German
	Research Foundation) -- Project-ID 279384907 -- SFB 1245. MRW
	acknowledges support from the Ministry of Science and Technology, Taiwan
	under Grant No.  107--2119--M--001--038, No. 108-2112-M-001-010, and the
	Physics Division of National Center for Theoretical Sciences.  The work
	of LMR is partly supported by the Spanish grant Nos PGC2018-094583-B-I00
	(MINECO). SAG and MRW thank the Yukawa Institute for Theoretical Physics
	in Kyoto for support in the framework of the YITP-T--18--06 workshop,
	during which several aspects of this work have been discussed.
	Computations were partly performed on the LOEWE-CSC computer managed by
	Center for Scientific Computing of the Goethe University Frankfurt.
\end{acknowledgments}

\bibliography{fission-rp}

\end{document}